\newif\ifarxiv
\arxivtrue

\ifarxiv

\documentclass[3p,times]{simplearticle}

\else

\documentclass[3p,times]{elsarticle}
\usepackage{ecrc}

\fi

\usepackage{epstopdf}

\ifarxiv
\else
\volume{00}

\firstpage{1}

\journalname{Nuclear Physics A}

\runauth{S. Aiola for the ALICE Coll.}

\jid{nupha}

\jnltitlelogo{Nuclear Physics A}

\fi

\usepackage{graphicx}
\usepackage{amsmath,amssymb}
\usepackage{comment}
\usepackage{subfig}
\usepackage[usenames]{color}
\usepackage{color}
\usepackage{float}

\usepackage{lineno}

\usepackage{hyperref}

\ifarxiv
\else
\biboptions{compress}
\fi

\newcommand{\com}[1]       {\relax}

\newcommand{\VZERO}        {\rm{VZERO}}

\newcommand{\PbPb}         {\mbox{Pb--Pb}}

\newcommand{\snn}          {\ensuremath{\sqrt{s_{\mathrm{NN}}}}}

\newcommand{\dpT}          {\ensuremath{\delta p_{\mathrm{T}}}}  
\newcommand{\pT}           {\ensuremath{p_{\mathrm{T}}}}

\newcommand{\RAA}          {\ensuremath{{R}_{\mathrm{AA}}}}

\newcommand{\kt}           {\ensuremath{k_{\mathrm{T}}}}

\newcommand{\GeVc}         {GeV/$c$}

\begin{document}

\ifarxiv
\else
\begin{frontmatter}
\fi


\title{Measurement of jet $p_{\rm T}$ spectra and $R_{\rm AA}$ in pp and Pb--Pb collisions at $\sqrt{s_{\rm NN}}$ = 2.76 TeV with the ALICE detector}

\ifarxiv
\author{Salvatore Aiola (for the ALICE Collaboration)}
\fi
\ead{salvatore.aiola@yale.edu}
\address{Yale University, New Haven, CT 06520}

\begin{abstract}
Hard-scattered partons provide an ideal probe for the study of the Quark-Gluon Plasma because they are produced prior to the formation 
of the QCD medium in heavy-ion collisions. 
Jet production is therefore susceptible to modifications induced by the presence of the medium (``jet quenching'').
Both RHIC and LHC experiments have provided compelling evidence of jet quenching. 
Jets are reconstructed in ALICE utilizing the central tracking system for the charged constituents 
and the Electromagnetic Calorimeter for the neutral constituents. Jet spectra are reported for central (0-10\%) and semi-central (10-30\%)
Pb--Pb events at $\snn=2.76$ TeV. The nuclear modification factor, determined using a pp baseline measured at the same collisional energy,
shows a strong suppression of jet production in central Pb--Pb collisions with the expected centrality ordering. Observations are in qualitative
agreement with medium-induced energy loss models. Furthermore, indication of a path-length dependence of 
jet suppression is inferred from measurements
of the yields relative to the orientation of the event plane.
\end{abstract}

\ifarxiv
\else

\end{frontmatter}
\fi


\section{Introduction}
\label{intro}
The study of jets in ultra-relativistic heavy-ion collisions is intimately connected 
to the investigation of the properties of the Quark-Gluon Plasma (QGP).
The QGP is a deconfined state of matter, in which the relevant degrees of freedom are those of strongly-interacting quarks and gluons.
Simple quantum-mechanical considerations, based on the Heisenberg principle of indetermination, 
imply that hard-scattering processes, with a large
momentum transfer, happen at a much smaller time scale as compared to that of the QGP formation, which is driven by many low momentum scatterings.
The subsequent transport of the hard scattered parton through the medium and its fragmentation are expected to be considerably modified through 
elastic collisions and medium-induced radiation~\cite{Baier:1995}. These phenomena are usually referred to as ``jet quenching''.
Measurements at both RHIC~\cite{STAR:2003c, PHENIX:2004a} 
and the LHC~\cite{CMS:2011c, ALICE:2014a, ATLAS:2013b}
have confirmed these predictions using a variety of observables.
Jet reconstruction can take advantage of the more accurate estimate of the energy of the parton compared
to high transverse momentum (\pT) single hadron measurements, which are often used as proxies for jets.

In these proceedings we report measurements of the jet nuclear modification factor and charged jet $v_2$ performed by the ALICE experiment
for \PbPb\ collisions at $\snn=2.76$~TeV. 
These results are based on data collected by ALICE in 2011 and extend previous measurements reported in Refs.~\cite{Aiola:2013, Reed:2013}.
\begin{figure}[t]
\centering
\subfloat[][Jet \pT{} spectra (0-10\% centrality class).]
{\includegraphics[width=.32\columnwidth]{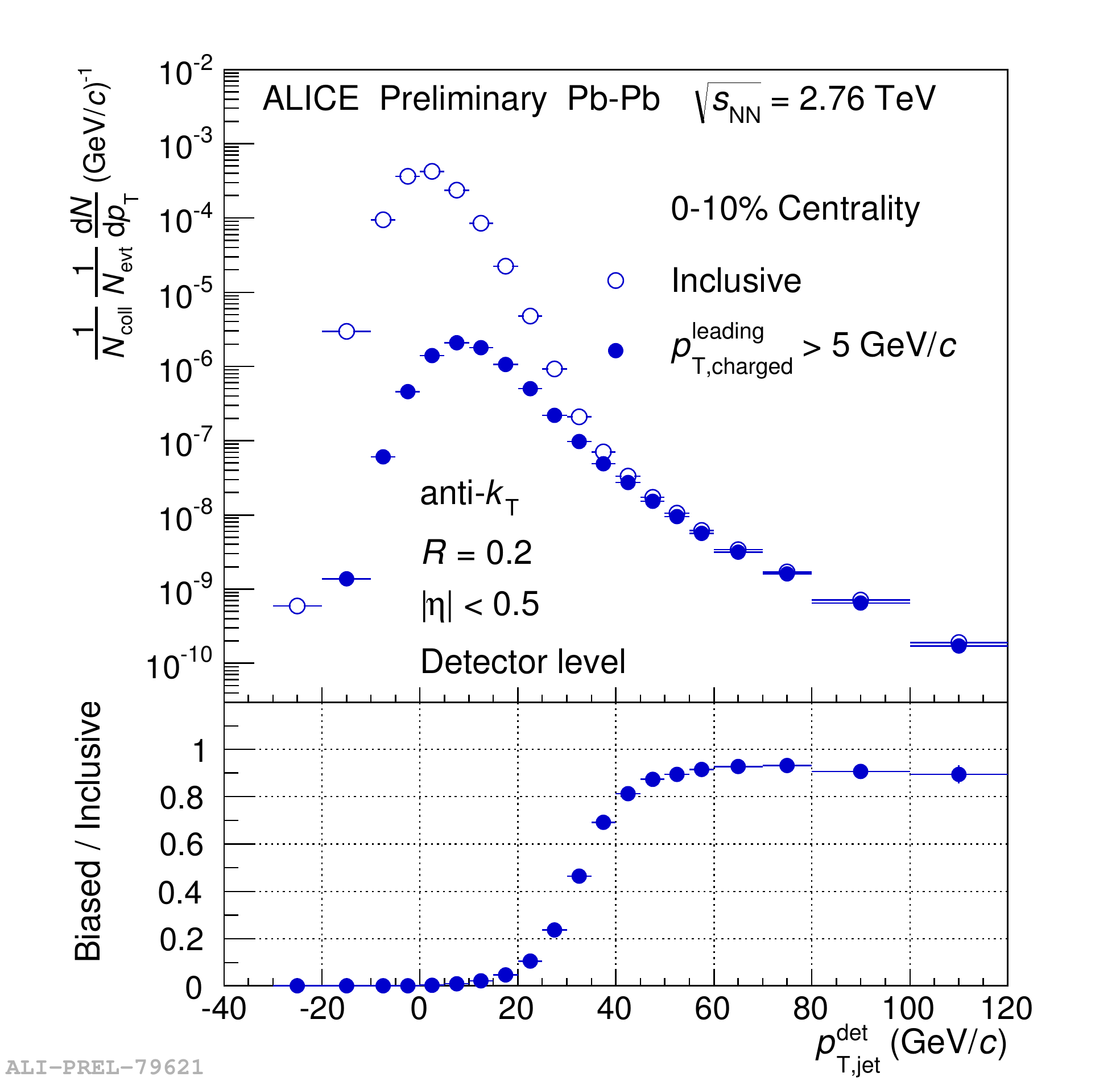}\label{fig:RawJetSpectraPt0}} \quad
\subfloat[][Jet \pT{} spectra (10-30\% centrality class).]
{\includegraphics[width=.32\columnwidth]{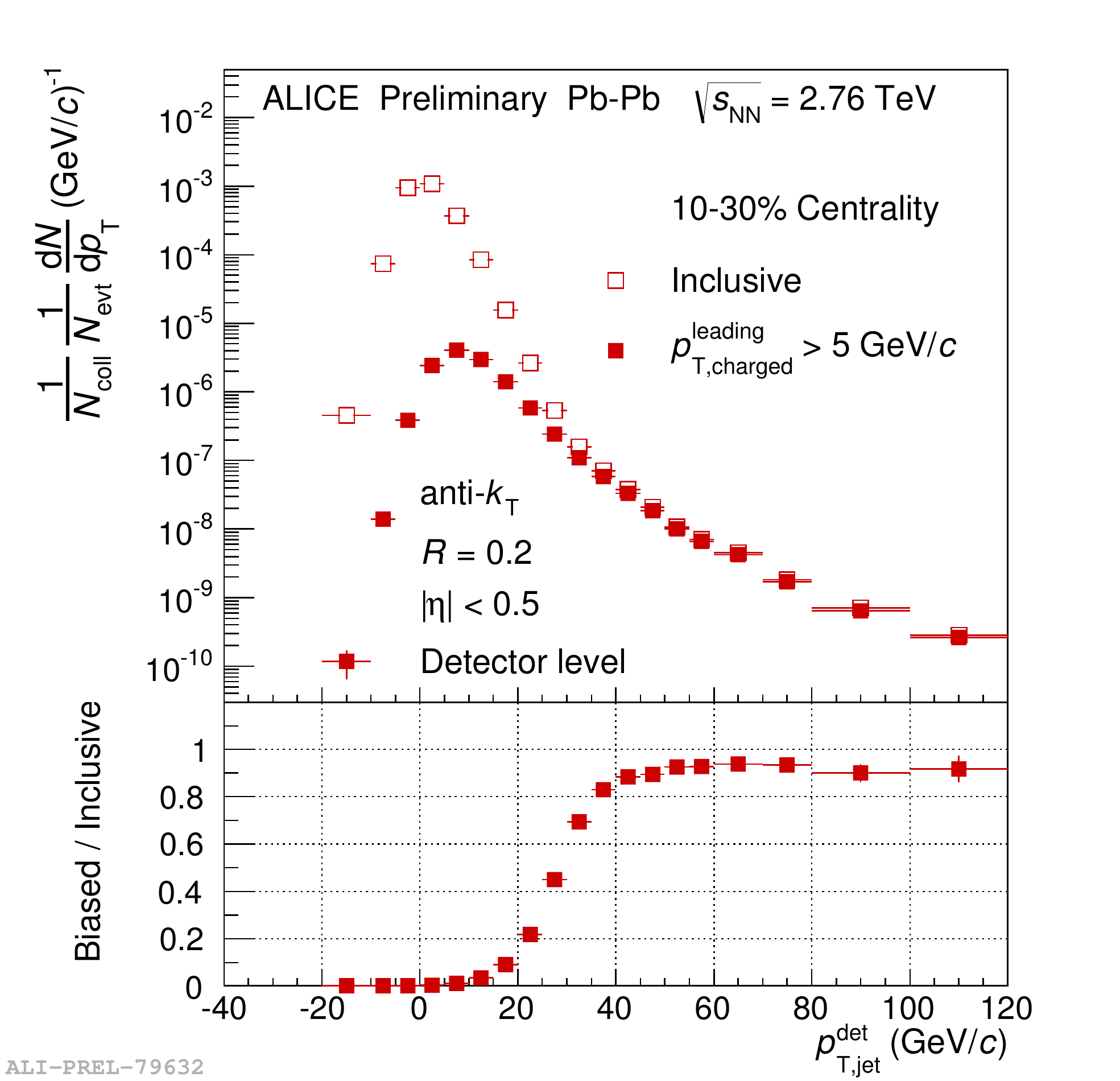}\label{fig:RawJetSpectraPt1}} \quad
\subfloat[][\dpT{} distributions.]
{\includegraphics[width=0.31\columnwidth]{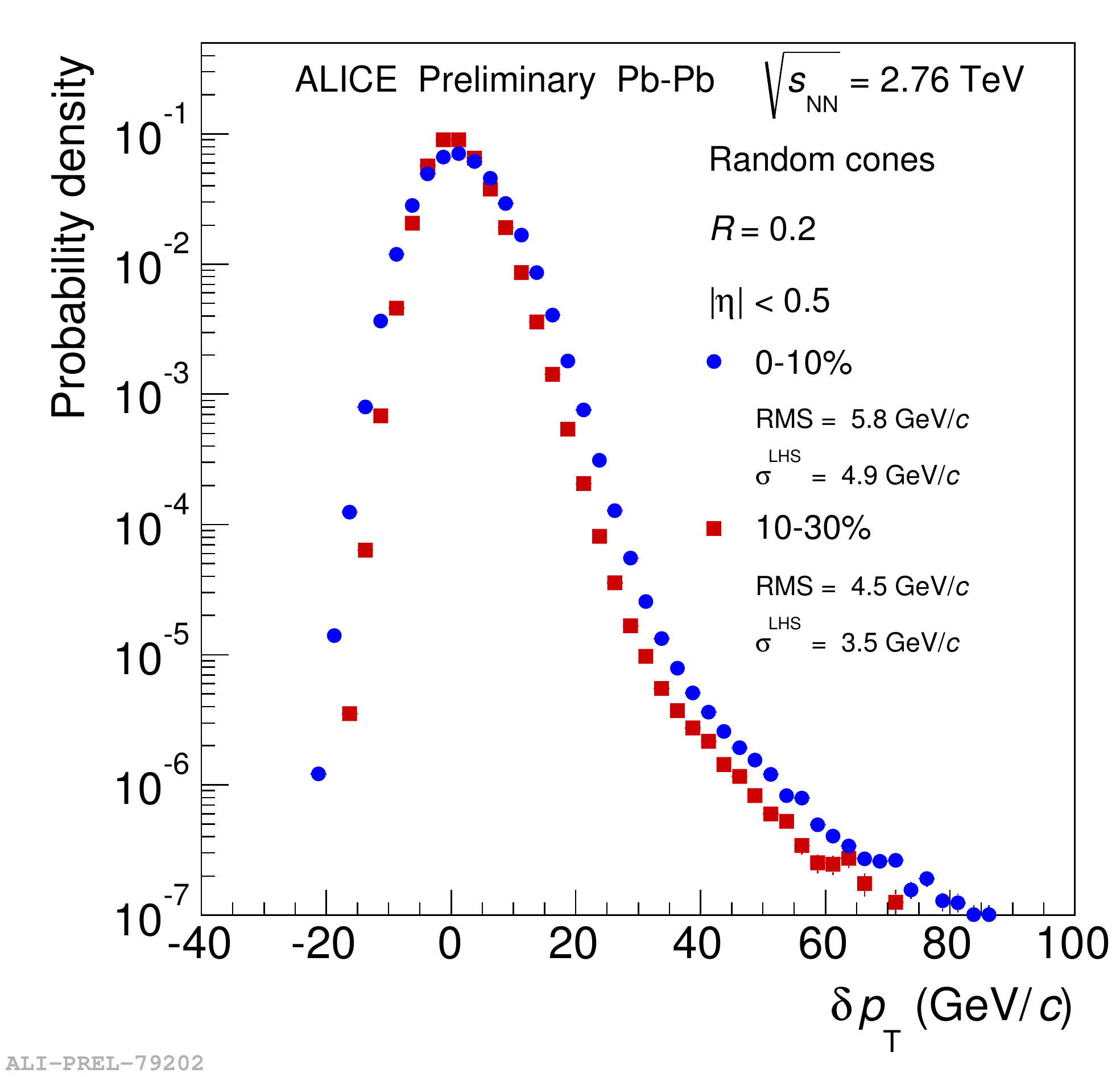}\label{fig:DeltaPt}}
\caption{Jet \pT{} spectra at detector level and \dpT{} distributions
in \mbox{Pb--Pb} collisions at $\snn=2.76$~TeV at mid-rapidity.
Jets are reconstructed using the anti-\kt{} algorithm 
with a resolution parameter $R=0.2$ in two centrality ranges: 0-10\% (left) and 10-30\% (middle).
Spectra are normalized by the number of events and the number of binary collisions obtained in a Glauber MC calculation~\cite{ALICE:2013b}.
Both the inclusive spectra and the spectra biased by requiring a leading hadron with $\pT>5$~\GeVc\ are shown.
The bottom panels show the ratios of the biased over the inclusive spectra. The \dpT{} distributions (right) 
are obtained using the random cone technique.}
\label{fig:RawJetSpectraDeltaPt}
\end{figure}
\section{Experimental setup and analysis techniques}
\label{sec:analysis}
For a complete description of the ALICE detector and its performance see Refs.~\cite{ALICE:2008} and \cite{ALICE:2014b}, respectively.
The main sub-detectors used in the present measurement are the \VZERO, the central tracking system and the Electromagnetic Calorimeter (EMCal).
The \VZERO\ detector consists of segmented scintillators covering the full azimuth at forward rapidity.
It is used to measure the centrality of the \PbPb\ events and also provides the set of minimum bias and centrality
triggers used to collect the presented data.
The ALICE tracking system consists of the Inner Tracking System (ITS), a six-layer silicon detector, and a large Time Projection Chamber (TPC).
The ITS provides a precise measurement of the first points of the tracks
and a precise determination of the primary vertex. The tracking system allows reconstruction of charged tracks ranging from very low momentum 
($\pT\approx 0.15$~GeV/$c$) to high momentum ($\pT\approx 100$~GeV/$c$), with good momentum resolution and tracking efficiency.
Tracks are reconstructed at mid-rapidity ($|\eta|<0.9$) and in full azimuth.
The EMCal is a Pb-scintillator sampling calorimeter, which covers mid-rapidity
($|\eta|<0.7$) and partial azimuth ($\Delta\phi=100^{\circ}$). In this analysis it is used
to measure the decay photons of the neutral mesons, that are not detected by the tracking system. Energy deposition
from charged particles, measured by the tracking system, is subtracted to avoid double counting their energy
in the jet reconstruction. The details of this correction are outlined in Ref.~\cite{ALICE:2014b}.

The methods utilized in this analysis follow closely those used for the charged jet suppression measurement reported in Ref.~\cite{ALICE:2014a} 
and for the pp jet cross section reported in Ref.~\cite{ALICE:2013c}.
The anti-\kt{} jet finding algorithm~\cite{Cacciari:2008c} with a resolution parameter $R=0.2$ has been employed in 
its FastJet~\cite{Cacciari:2012} implementation. Jets are required to be fully contained within the EMCal acceptance.
Following the proposal in Ref.~\cite{Cacciari:2008b}, the average background, $\rho$, is calculated, event-by-event, as the median of the \pT{}
density (jet \pT{} over jet area) of the jets reconstructed by the \kt{} algorithm.
The average background is subtracted jet-by-jet: $p_{\rm T,jet}^{\rm reco}=p_{\rm T,jet}^{\rm raw}-\rho \times A_{\rm jet}$,
where $p_{\rm T,jet}^{\rm raw}$ and $A_{\rm jet}$ are respectively the transverse momentum and the area of the anti-\kt\ jet candidate. 
Figures~\ref{fig:RawJetSpectraPt0} and \ref{fig:RawJetSpectraPt1} show the jet \pT{} spectra at detector level, 
after the subtraction of the average background.
Both the inclusive spectra and the spectra biased by requiring a leading hadron $\pT>5$ \GeVc\ are shown.
The bias is applied in order to suppress the combinatorial background.
\begin{figure}[tb]
\centering
\subfloat[][Jet \pT\ spectra.]
{
\includegraphics[width=.35\columnwidth]{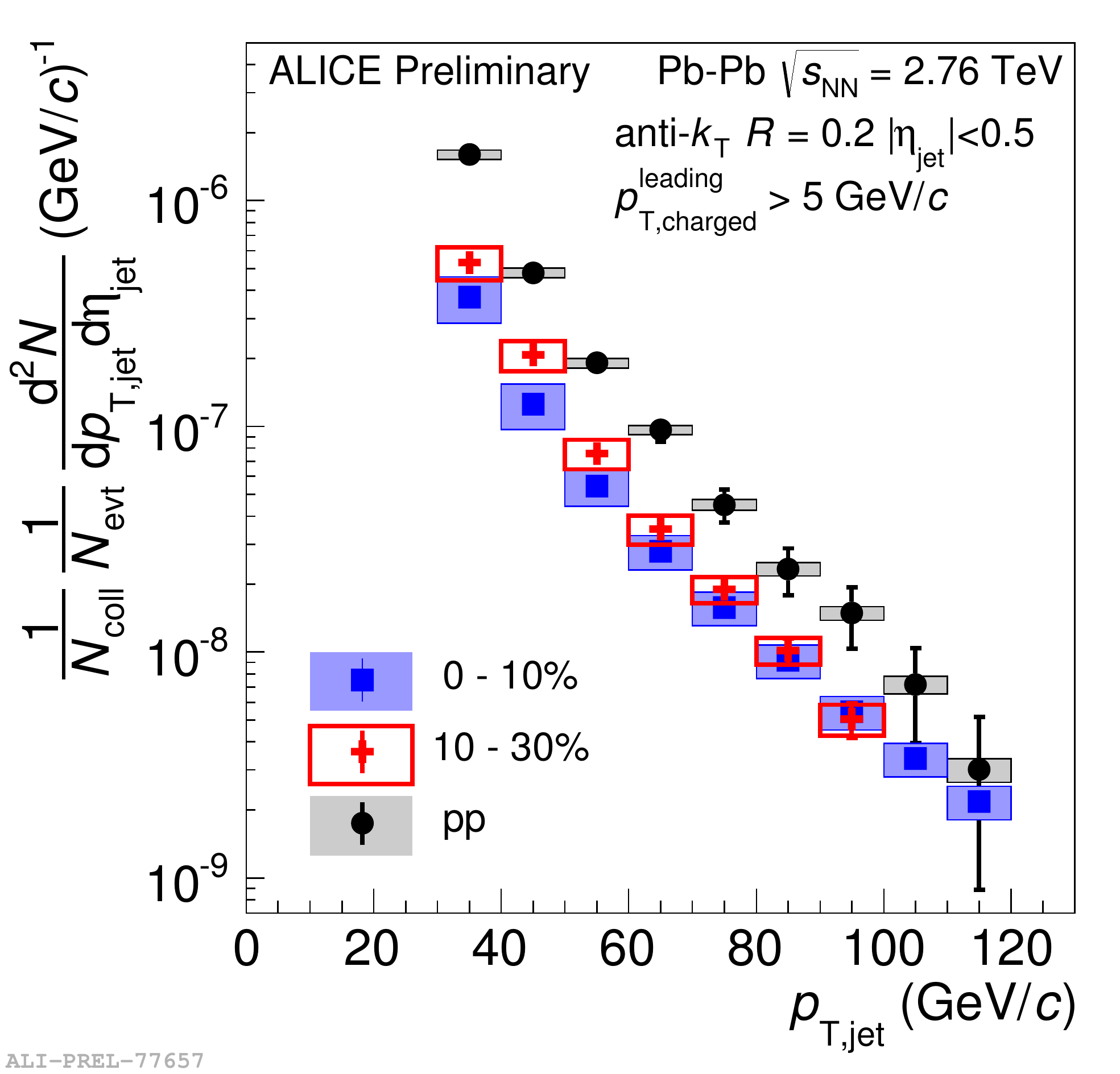}
\label{fig:Spectra}
} \quad
\subfloat[][Nuclear modification factor.]
{
\includegraphics[width=.35\columnwidth]{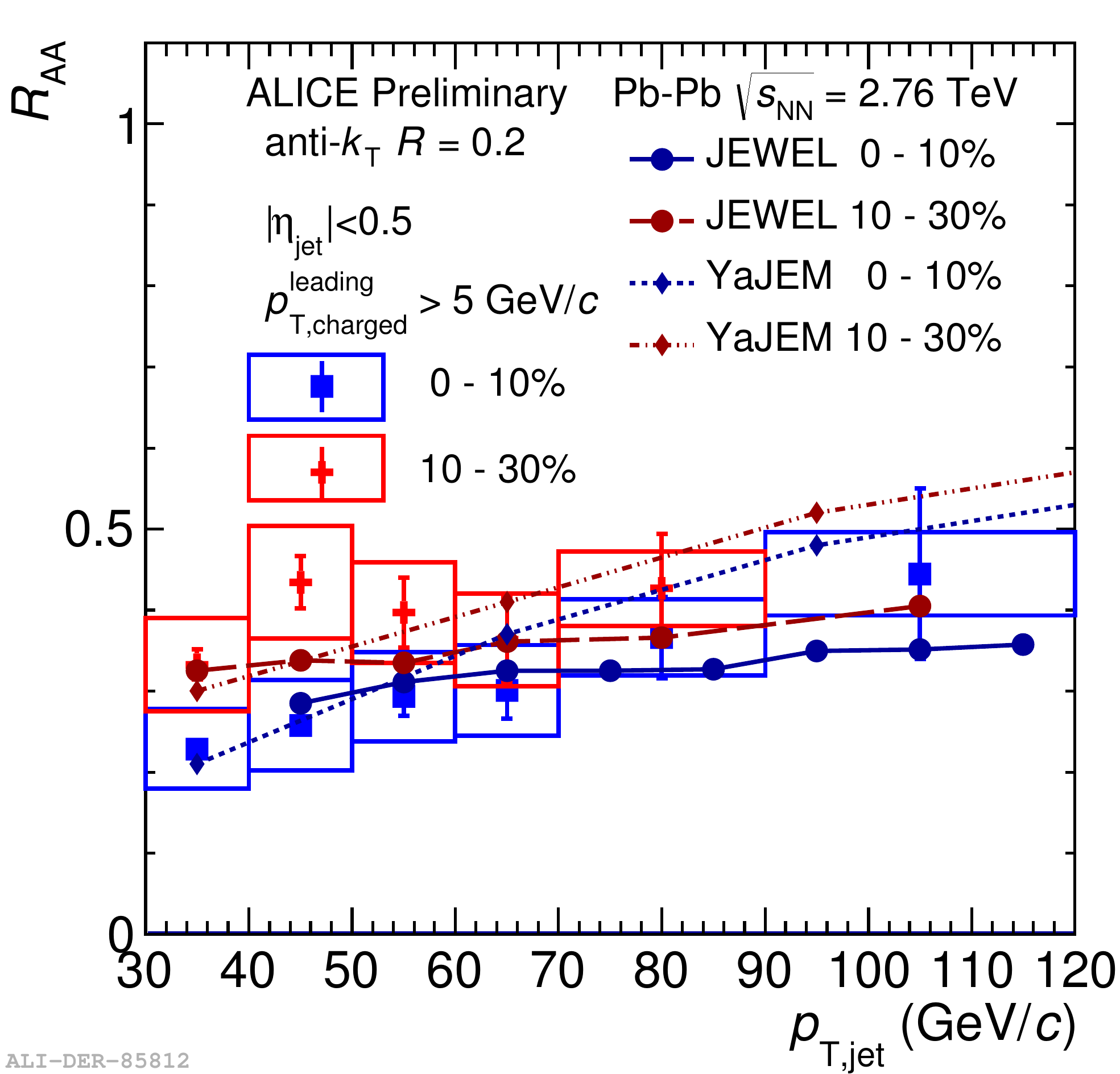}
\label{fig:Raa}
} 
\caption{Jet \pT\ spectra (left) and nuclear modification factor (right) at mid-rapidity for 0-10\% and 10-30\% centrality class \PbPb\ events 
at \snn$=2.76$ TeV (color online). The measured \RAA\ is compared with two model predictions, YaJEM~\cite{Renk:2013} 
and JEWEL~\cite{Zapp:2014}, see text for details.}
\label{fig:SpectraRaa}
\end{figure}
Background fluctuations are estimated using the random cone technique~\cite{ALICE:2012a}.  
The root-mean-square widths of the \dpT\ distributions, shown in Fig.~\ref{fig:DeltaPt},
are $5.8$ \GeVc\ for the 0-10\% centrality class and $4.5$ \GeVc\ for the 10-30\%. 
The response of the detector to jets has been quantified through a simulation which makes use of the PYTHIA6 event generator~\cite{Sjostrand:2006} 
and the GEANT3 transport code~\cite{Brun:1994}. 
The results reported in Section~\ref{sec:results} are obtained after correcting for both detector response and background fluctuations using 
standard regularized unfolding methods~\cite{Hocker:1995, Dagostini:1995}.
\section{Results}
\label{sec:results}
Figure~\ref{fig:Spectra} shows the anti-\kt\ $R=0.2$ jet \pT\ spectra at mid-rapidity for the 0-10\% and the 10-30\% centrality classes. 
In order to compare with the same measurement performed in pp
collisions at the same \snn~\cite{ALICE:2013c}, the jet yield in \PbPb\ has been divided by the number of binary collisions
calculated in a Monte Carlo Glauber model~\cite{ALICE:2013b} that assumes independent binary nucleon-nucleon collisions.
The systematic uncertainty is dominated by the tracking efficiency uncertainty and the unfolding regularization and is \pT\ dependent, 
with a value of
about 18\% at $p_{\rm T,jet}=60$ \GeVc\ for the central events and only slightly smaller for the semi-central events. 

The nuclear modification factor \RAA\ is defined as the ratio of the \PbPb\ per-event yield over the pp cross section~\cite{ALICE:2013c} 
multiplied by $T_{\mathrm{AA}}=N_{\mathrm{coll}} / \sigma^{\mathrm{inel}}_{\mathrm{pp}}$, obtained in the same
Glauber calculation mentioned above. The measured \RAA\ is shown in Fig.~\ref{fig:Raa}. A strong suppression, with the expected centrality
ordering, is observed and is in qualitative agreement with two 
energy loss model predictions, superimposed on the data. 
Both models use a combination of Glauber MC, perturbative QCD (pQCD) Leading Order (LO) calculations and PYTHIA,
to describe the initial state, the hard scattering and the hadronization into final-state colorless particles. 
They both implement a hydrodynamic description of the medium.
They differ in the specific way in which the interaction between the shower parton and the medium is modeled.
In YaJEM~\cite{Renk:2013} the energy loss mechanism is implemented through a pair of transport coefficients, 
one representing the increase of the parton's virtuality in the medium (radiative energy loss), and 
the other accounting for the collisional energy loss; in JEWEL~\cite{Zapp:2014} an average over a microscopic description
of the single parton-parton scatterings is implemented, using a MC model for the LPM interference.
\begin{figure}[t]
\centering
\subfloat[][0-5\% centrality class.]
{\includegraphics[width=.4\columnwidth]{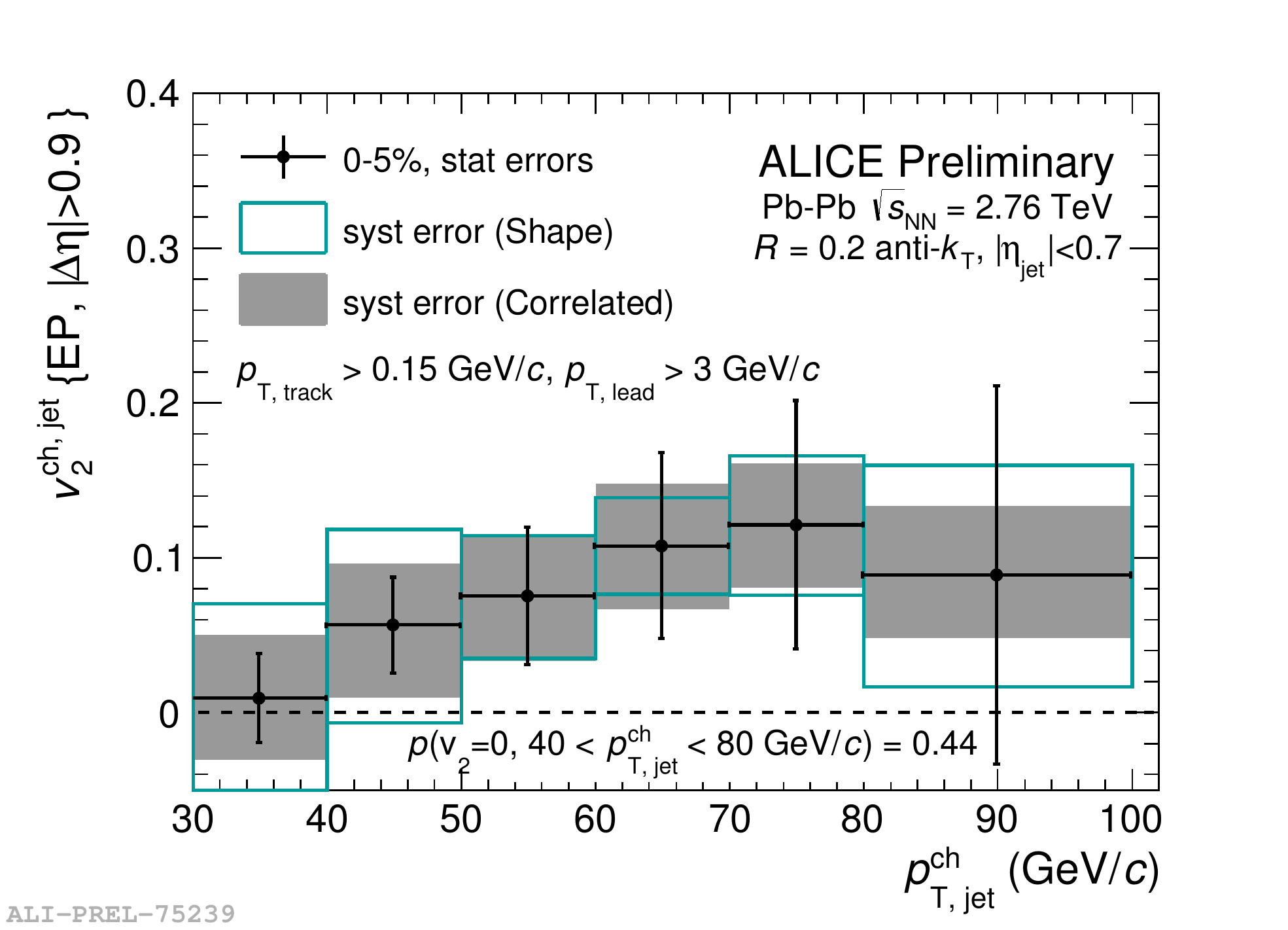}} \quad
\subfloat[][30-50\% centrality class.]
{\includegraphics[width=.4\columnwidth]{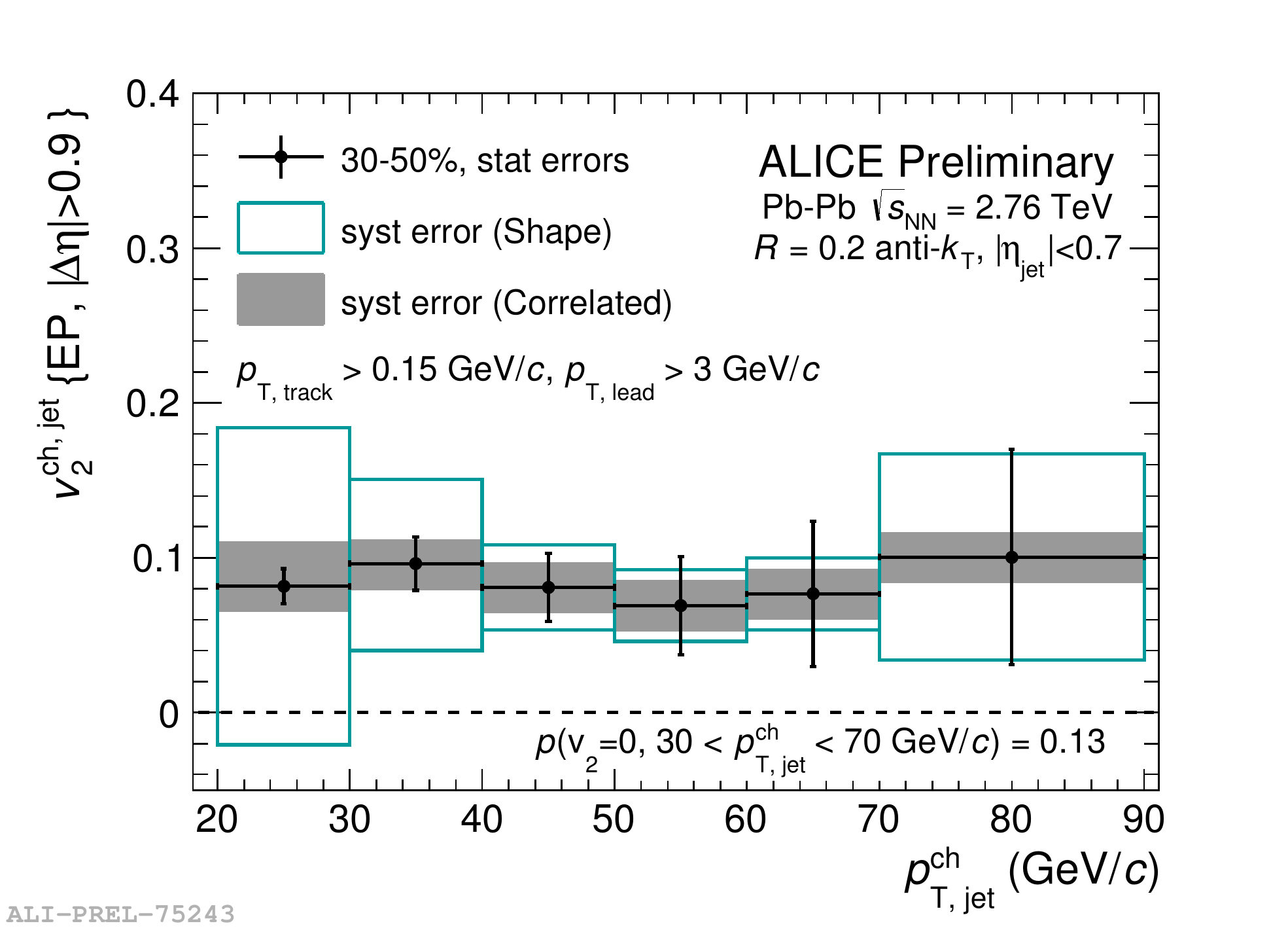}} 
\caption{Charged jet $v_2$ in \mbox{Pb--Pb} collisions at $\snn=2.76$~TeV at mid-rapidity. Jets are required to have a leading
hadron with $\pT>3$ \GeVc.}
\label{fig:Jetv2}
\end{figure}

The path-length dependence of the jet energy loss has been investigated by studying the charged jet yield with respect to the 
event plane orientation. The QGP formed in semi-central collisions is expected to take an ellipsoidal shape elongated in the direction
perpendicular to the reaction plane, 
which has been confirmed by the measurement
of the azimuthal anisotropy of low-momentum particle production~\cite{ALICE:2010b}. 
High $Q^2$ partons produced out-of-plane are expected to be more heavily modified because
they travel a longer path in the medium. The jet $v_2$ is defined as the second coefficient 
of the Fourier expansion of the jet yield with respect
to the azimuthal angle between the jet axis and the event plane.
For this measurement,
the average background and the background fluctuations have been measured and subtracted differentially with respect to 
the angle with the event plane. The observed $v_2$ has been corrected for the finite event plane resolution
by using a three sub-event technique~\cite{Poskanzer:1998}. Fig.~\ref{fig:Jetv2} shows the charged jet $v_2$
for central (0-5\%) and semi-central (30-50\%) \PbPb\ events. 
In estimating the systematic uncertainties, the correlations between the in-plane and out-of-plane measurements have been taken into account.
While we do not observe a significant indication for a non-zero jet $v_2$ in central events, the semi-central events shows 
a hint of a difference between the in- and out-of-plane nuclear modification factor, 
although the result is also still compatible with $v_2=0$ for jets.
\section{Conclusions}
In these proceedings we have reported new measurements of jet suppression performed by the ALICE experiment.
The measured suppression for both central and semi-central events is in qualitative agreement with both the YaJEM and JEWEL models.
In order to obtain a tighter constraint on the models the path-length dependence of the parton energy loss has been
explored by measuring the charged jet $v_2$, which shows hints of such an effect for semi-central events.
ALICE has also measured jet yields in p--Pb collisions~\cite{Haake:2013, Connors:2014}, which show that cold nuclear matter effects in
the initial state cannot account for the observed suppression in \PbPb\ collisions. These observations support the
interpretation of the observed suppression as a medium-induced modification of the final state parton shower.
\ifarxiv
\bibliographystyle{elsarticle-num}
\section*{References}
{\footnotesize\bibliography{biblio}}

\begin{thebibliography}{10}
\expandafter\ifx\csname url\endcsname\relax
  \def\url#1{\texttt{#1}}\fi
\expandafter\ifx\csname urlprefix\endcsname\relax\def\urlprefix{URL }\fi
\expandafter\ifx\csname href\endcsname\relax
  \def\href#1#2{#2} \def\path#1{#1}\fi

\bibitem{Baier:1995}
R.~Baier, Y.~L. Dokshitzer, S.~Peigne, D.~Schiff, Phys.Lett. B345 (1995)
  277--286  \href {http://dx.doi.org/10.1016/0370-2693(94)01617-L}
  {\path{doi:10.1016/0370-2693(94)01617-L}}.

\bibitem{STAR:2003c}
{\bfseries STAR} Collaboration, C.~Adler, et~al., Phys.Rev.Lett. 90 (2003)
  082302  \href {http://dx.doi.org/10.1103/PhysRevLett.90.082302}
  {\path{doi:10.1103/PhysRevLett.90.082302}}.

\bibitem{PHENIX:2004a}
{\bfseries PHENIX} Collaboration, S.~S. Adler, et~al., Phys. Rev. C69 (2004)
  034910  \href {http://dx.doi.org/10.1103/PhysRevC.69.034910}
  {\path{doi:10.1103/PhysRevC.69.034910}}.

\bibitem{CMS:2011c}
{\bfseries CMS} Collaboration, S.~Chatrchyan, et~al., Phys. Rev. C 84 (2011)
  024906  \href {http://dx.doi.org/10.1103/PhysRevC.84.024906}
  {\path{doi:10.1103/PhysRevC.84.024906}}.

\bibitem{ALICE:2014a}
{\bfseries ALICE} Collaboration, B.~Abelev, et~al., JHEP 1403 (2014) 013  \href
  {http://arxiv.org/abs/1311.0633} {\path{arXiv:1311.0633}}, \href
  {http://dx.doi.org/10.1007/JHEP03(2014)013}
  {\path{doi:10.1007/JHEP03(2014)013}}.

\bibitem{ATLAS:2013b}
{\bfseries ATLAS} Collaboration, G.~Aad, et~al., Phys.Lett. B719 (2013)
  220--241  \href {http://arxiv.org/abs/1208.1967} {\path{arXiv:1208.1967}},
  \href {http://dx.doi.org/10.1016/j.physletb.2013.01.024}
  {\path{doi:10.1016/j.physletb.2013.01.024}}.

\bibitem{Aiola:2013}
S.~Aiola {(for the ALICE Collaboration)},
  \href{http://stacks.iop.org/1742-6596/446/i=1/a=012005}{Journal of Physics:
  Conference Series} 446 (2013) 012005 \href {http://arxiv.org/abs/1304.6668}
  {\path{arXiv:1304.6668}}.

\bibitem{Reed:2013}
R.~Reed {(for the ALICE Collaboration)},
  \href{http://stacks.iop.org/1742-6596/446/i=1/a=012006}{Journal of Physics:
  Conference Series} 446 (2013) 012006 \href {http://arxiv.org/abs/1304.5945}
  {\path{arXiv:1304.5945}}.

\bibitem{ALICE:2013b}
{\bfseries ALICE} Collaboration, B.~Abelev, et~al., Phys.Rev. C88~(4) (2013)
  044909  \href {http://arxiv.org/abs/1301.4361} {\path{arXiv:1301.4361}},
  \href {http://dx.doi.org/10.1103/PhysRevC.88.044909}
  {\path{doi:10.1103/PhysRevC.88.044909}}.

\bibitem{ALICE:2008}
{\bfseries ALICE} Collaboration, K.~Aamodt, other, Journal of Instrumentation 3
  (2008) S08002  .

\bibitem{ALICE:2014b}
{\bfseries ALICE} Collaboration, B.~B. Abelev, et~al.  \href
  {http://arxiv.org/abs/1402.4476} {\path{arXiv:1402.4476}}.

\bibitem{ALICE:2013c}
{\bfseries ALICE} Collaboration, B.~Abelev, et~al., Phys.Lett. B722 (2013)
  262--272  \href {http://arxiv.org/abs/1301.3475} {\path{arXiv:1301.3475}},
  \href {http://dx.doi.org/10.1016/j.physletb.2013.04.026}
  {\path{doi:10.1016/j.physletb.2013.04.026}}.

\bibitem{Cacciari:2008c}
M.~Cacciari, G.~Salam, G.~Soyez, JHEP 04~(063) (2008) 063  .

\bibitem{Cacciari:2012}
M.~Cacciari, G.~P. Salam, G.~Soyez, Eur.Phys.J. C72 (2012) 1896  \href
  {http://arxiv.org/abs/1111.6097} {\path{arXiv:1111.6097}}, \href
  {http://dx.doi.org/10.1140/epjc/s10052-012-1896-2}
  {\path{doi:10.1140/epjc/s10052-012-1896-2}}.

\bibitem{Cacciari:2008b}
M.~Cacciari, G.~Salam, Physics Letters B 659~(1-2) (2008) 119--126  .

\bibitem{Renk:2013}
T.~Renk, Phys.Rev. C88~(1) (2013) 014905  \href
  {http://arxiv.org/abs/1302.3710} {\path{arXiv:1302.3710}}, \href
  {http://dx.doi.org/10.1103/PhysRevC.88.014905}
  {\path{doi:10.1103/PhysRevC.88.014905}}.

\bibitem{Zapp:2014}
K.~C. Zapp, Physics Letters B 735 (2014) 157 -- 163  \href
  {http://arxiv.org/abs/1312.5536} {\path{arXiv:1312.5536}}, \href
  {http://dx.doi.org/10.1016/j.physletb.2014.06.020}
  {\path{doi:10.1016/j.physletb.2014.06.020}}.

\bibitem{ALICE:2012a}
{\bfseries ALICE} Collaboration, B.~Abelev, et~al., JHEP 1203~(053) (2012) 053
  .

\bibitem{Sjostrand:2006}
T.~Sj{\"o}strand, S.~Mrenna, P.~Skands, Journal of High Energy Physics 5~(26)
  (2006) 1--581  .

\bibitem{Brun:1994}
R.~Brun, F.~Carminati, S.~Giani, CERN Program Library Long Write-up, W5013  .

\bibitem{Hocker:1995}
A.~Hocker, V.~Kartvelishvili, NIM A372 (1996) 469--481  \href
  {http://arxiv.org/abs/hep-ph/9509307} {\path{arXiv:hep-ph/9509307}}, \href
  {http://dx.doi.org/10.1016/0168-9002(95)01478-0}
  {\path{doi:10.1016/0168-9002(95)01478-0}}.

\bibitem{Dagostini:1995}
G.~D'Agostini, NIM 362 (1995) 487  .

\bibitem{ALICE:2010b}
{\bfseries ALICE} Collaboration, K.~Aamodt, et~al., Phys. Rev. Lett. 105 (2010)
  252302  \href {http://dx.doi.org/10.1103/PhysRevLett.105.252302}
  {\path{doi:10.1103/PhysRevLett.105.252302}}.

\bibitem{Poskanzer:1998}
A.~M. Poskanzer, S.~A. Voloshin,
  \href{http://link.aps.org/doi/10.1103/PhysRevC.58.1671}{Phys. Rev. C} 58
  (1998) 1671--1678  \href {http://dx.doi.org/10.1103/PhysRevC.58.1671}
  {\path{doi:10.1103/PhysRevC.58.1671}}.

\bibitem{Haake:2013}
R.~Haake {(for the ALICE Collaboration)}, PoS EPS-HEP2013 (2014) 176 \href
  {http://arxiv.org/abs/1310.3612} {\path{arXiv:1310.3612}}.

\bibitem{Connors:2014}
M.~Connors {(for the ALICE Collaboration)},
  \href{http://www.sciencedirect.com/science/article/pii/S0375947414004746}{Nuclear
  Physics A} 931 (2014) 1174 \href {http://arxiv.org/abs/1409.3468}
  {\path{arXiv:1409.3468}}, \href
  {http://dx.doi.org/10.1016/j.nuclphysa.2014.09.092}
  {\path{doi:10.1016/j.nuclphysa.2014.09.092}}.

\end{thebibliography}
\fi

\end{document}